\documentstyle[prl,aps,twocolumn]{revtex}
\input{epsf}
\begin{document}
\bibliographystyle{pnas}
\renewcommand{\thefootnote}{\fnsymbol{footnote}}
\title{Two State Behavior in a Solvable Model of $\beta$-hairpin folding}
\author{Chinlin Guo$^*$, Herbert Levine$^{*,\ddag }$, and 
	David A. Kessler$^\dag$}
\address{
 $^*$Department of Physics, University of California, San Diego, 
9500 Gilman Drive, La Jolla, CA 92093-0319 \\
 $^\dag$Department of Physics, Bar-Ilan University, Ramat-Gan, Israel}
\author{$\ddag$Author to whom correspondence should be addressed. \\ 
Email: levine@herbie.ucsd.edu.\\ 
Phone: (858)-534-4844. Fax: (858)-534-7697.}
\date{\today}
\maketitle
{\bf
Understanding the mechanism of protein secondary structure formation is
an essential part of protein-folding puzzle. Here we describe a simple
model for the formation of the $\beta$-hairpin, motivated by the fact that
folding of a $\beta$-hairpin captures much of the basic physics of protein
folding. We argue that the coupling of ``primary'' backbone stiffness and
``secondary'' contact formation (similar to the coupling between the 
``secondary'' and ``tertiary'' structure in globular proteins), caused for
example by side-chain packing regularities, is responsible for producing an 
all-or-none 2-state $\beta$-hairpin formation. We also
develop a recursive relation to compute the phase diagram and single 
exponential folding/unfolding rate arising via a dominant transition state.
}

\section{Introduction}

Recent years have seen an intense effort aimed at elucidating the
physics underlying protein folding \cite{reviews}. One crucial question 
concerns the nature of the transition from the random coil to the native
conformation. Essentially, we wish to discover the critical
parameters governing whether this transition is first-order (``all or
none'') or continuous and furthermore we wish to characterize the transition
kinetics. In this paper, we focus on these issues in a
very simple context, the formation of a $\beta$-hairpin. 

For the past forty years, the $\alpha$ helix-coil transition has been 
extensively studied \cite{Poland}. Here, the transition is in general 
continuous rather than abrupt; hence there is no 2-state behavior. 
In comparison, however, a recent experiment \cite{Munoz2} has shown that
 $\beta$-hairpin formation can exhibit a 2-state collective behavior
between the random coil (unfolded) and native hairpin (folded) states. Recent 
computational studies \cite{Klimov,Kolinski} have concluded that 
hydrogen-bond formation between the two sides of the hairpin is 
insufficient to produce an all-or-none 2-state behavior. Instead,
one must also take into account the hydrophobic side chain packing 
regularities. Translated into the language of simple models, one
would therefore expect that a simple pairwise $\rm G\bar{o}$-like interaction
would not give rise to an all-or-none transition; instead, one must add
additional terms corresponding to coupling the ``secondary'' structure
(the contact formation between residues on the opposite sides 
of the hairpin) with the ``primary'' structure (the stiffness of the backbone).

The purpose of this paper is to introduce a simple, exactly solvable
model which allows one to calculate the equilibrium states and
the transition kinetics of a model with this type of coupling. 
We describe the hairpin by two Gaussian chains (attached at
the turn of the hairpin) whose interaction is described by two types of terms. 
There is a pairwise $\rm G\bar{o}$-like interaction mimicking the
hydrogen-bond formation and a short ranged many-body interaction 
approximating the side chain packing regularities. To simplify our model, 
we assume that the hydrogen bond formation and side chain
packing regularities are uniformly distributed among the  
residues. This allows us to develop a set of recursion relations
for the exact determination of the partition function, and show the range 
of parameters for which the hairpin has 2-state behavior. Finally, we can
estimate the (single exponential) folding/unfolding rate via calculating 
the thermodynamic weight of the ``critical'' droplet/bubble.

\section{The Model Hamiltonian}

\begin{figure}
\centerline 
{\epsfxsize = 8.0cm \epsffile{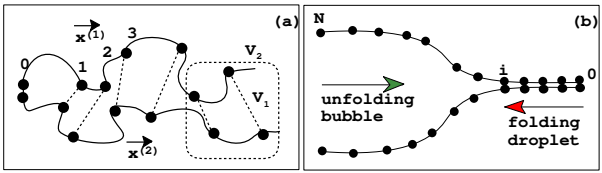}}
\vspace{.2cm}
\caption{ 
(a) A $\beta$ hairpin polymer with two interacting
branches. Each branch contains $N$ residues with $\rm G\bar{o}$-like 
inter-chain hydrogen bond interactions ($V_1$, the dashed line) and 
with cooperative side chain packing terms ($V_2$, the dashed box).
Here only the $\rm G\bar{o}$-interacting residues are shown. The degrees of 
freedom of those non-$\rm G\bar{o}$-interacting residues are assumed to be 
``renormalized'' into those $\rm G\bar{o}$-interacting ones.  
(b) The hairpin zippered from the $\beta$ turn.
In the folding (unfolding) regime, the droplet (bubble) will expand.
}\label{h1}
\end{figure}

We consider a hairpin polymer composed of two interacting Gaussian
chains (labeled as branch 1 and 2) connected by a $\beta$
turn at the proximal end, labeled as 
sequence index $i=0$ in fig.\ref{h1}(a). To
have a unique native structure, we impose $N$-pairwise $\rm G\bar{o}$
interactions on this polymer, which mimic the hydrogen bonds formed
by the 2$N$ residues. Our approach assumes that one can write
down an effective Hamiltonian in terms of the spatial coordinates
$\vec{x}_i^{(k)}$ ($i$ is residue index counted 
away from the $\beta$ turn and the superscript $k=1,2$ stands for branch 
labeling) of these $\rm G\bar{o}$ interacting residues. The
non-interacting part of this Hamiltonian is simply  
$\sum_{k=1,2}H_{\rm Gau}^{(k)} +H_0$ with $H_0 = \kappa 
|\vec{x}_0^{(1)}-\vec{x}_0^{(2)}|^2$ and
$$ H_{\rm Gau}^{(k)} =\kappa \sum_{i=1}^N |\nabla\vec{x}_i^{(k)}|^2$$ 
Here $\nabla\vec{x}_i^{(k)}=\vec{x}_i^{(k)}-\vec{x}_{i-1}^{(k)}$ is the
vector connected nearest-neighbor residues on each chain, and
$\kappa$ is backbone stiffness. 
 
The second ingredient of the Hamiltonian is the inter-chain interaction.
As already discussed, we use the $\rm G\bar{o}$ 
interaction to mimic the hydrogen bond formation. These bonds result
from a short-distance proximity between the donor and acceptor residues 
(such that the water or ion molecule serving as counter-ion shielding can 
be squeezed out).
Having a solvable model requires us to assume that the
binding strength is uniformly distributed along the chain. This
leads to the form 
 $V_{\rm Hb}(|\vec{x}_i|)=-V_1\Delta(|\vec{x}_i|)\equiv -V_1\Delta_i$ 
with $V_1\ge0$, where $\Delta_i=1$ if the inter-residue distance 
 $|\vec{x}_i| \equiv |\vec{x}^{(1)}-\vec{x}^{(2)}|$ falls into an 
effective attraction window $|\vec{x}_i|\le r_0$ and 0 otherwise. 
This "box" approximation to the potential has also 
been used by other groups \cite{Cule,Pande,Takada}.
\global\firstfigfalse

The final effects we consider arise from the (hydrophobic) side chain
packing effect. This interaction depends both on the formation
of inter-chain contacts and on the alignment of the 
local backbone \cite{Kolinski,Munoz1}. In principle, there are two
separate pieces that one can add to our model to mimic this interaction.
First, the presence of one hydrogen bond might, via the
local structural dynamics (such as squeezing out water molecules), cooperatively help
other neighboring residues form contacts. 
Such a collective term has the form 
 $H_{\rm Sc}=-V_2 \sum_{ij}\Delta_i \Delta_j J_{ij}$; as always, $\Delta_i$ 
indicates the contact formation of residue pair $i$, and in
addition $J_{ij}$ is a coupling function indicating 
the range of cooperativity. The simplest short-range assumption is
that $J_{ij}$ is 1 when $i,j$ are nearest-neighbors and 0 otherwise;
one might also imagine a longer-ranged hierarchical side packing
scheme such as that considered by Hansen et al.\cite{Hansen}; here we stick to 
the simplest possibility. This new term reflects a coupling from the primary 
structure to the secondary one.
In addition, forming a contact limits the conformations available to the 
chain; in our Gaussian model, this corresponds to an increase in the chain 
stiffness between $i$ and $i+1$ via a term 
 $\kappa\gamma(\Delta_i+\Delta_{i+1})$. 

It is straightforward to integrate out the mean coordinate
 $\vec{X}_i=[\vec{x}^{(1)}+\vec{x}^{(2)}]/2$ and express the
Hamiltonian in terms of the inter-residue distances only. 
The remaining Gaussian component will be
\begin{equation}\label{Gaussian}
H_{\rm Gau}={\kappa\over2}\left[ \sum_{k=1}^N | \nabla\vec{x}_k|^2
+2 |\vec{x}_0^{(1)}-\vec{x}_0^{(2)}|^2\right]
\end{equation} 
Without loss of the generality, we assume that
the $\beta$ turn is fixed at $\vec{x}_0=0$ and drop the Gaussian component
$|\vec{x}_0^{(1)}-\vec{x}_0^{(2)}|^2$. This should not affect our result 
significantly if the system size is large enough. 
The interaction component is 
\begin{eqnarray}\label{int}
H_{\rm int} 
&=&{\gamma \kappa\over2}\sum_{k=2}^N
|\nabla\vec{x}_k|^2(\Delta _k + \Delta_{k-1}) \nonumber\\
&&\hspace{.1cm}
-\sum_{k=1}^N\Delta_k \left[ V_1+V_2 \Delta_{k-1}\right]
\end{eqnarray}
 
We now proceed to find the partition function for this model.
Since we are interested in the case of short-ranged $\rm G\bar{o}$ 
interaction, (i.e., the effective contact distance $r_0$  
is far smaller than the thermodynamic mean inter-residue distance 
 $1/\sqrt{\beta\kappa}$), it is 
reasonable to replace the Gaussian connectivity term
 ${\kappa\over2}|\nabla\vec{x}_k|^2$ by
 ${\kappa\over2}|\vec{x}_k|^2$ if $\vec{x}_{k-1}$ is in a contact 
position. This approximation will greatly simplify the calculation.

\section{The phase diagram}
 
Given the above simplification, our model can be exactly solved in terms
of a set of recursion relations. We are eventually interested in 
the full partition function, which reads
\begin{equation}\label{Z1}
Z_N \ = \ \prod_{k=1}^N\int  
\left[{\beta\kappa\over2\pi}\right]^{d\over2} d^dx_k
e^{-\beta H(\{\vec{x}\})}\Biggr|_{\vec{x}_0=0}
\end{equation} 
where $d$ is the dimensionality which we will take as 3. 
It will also be convenient to consider several ``restricted'' partition
functions, corresponding to summing over all the states consistent with
some extra constraints.  First, we define $Z_{N,n_f}$ as 
the restricted partition function of an ensemble that contains all 
configurations specified by a particular number of contacts number $n_f$. The 
full partition function is just a sum of  $Z_{N,n_f}$ over this contact number.
Second, we define $Z_{N,n_f}^c$ as the restricted partition function of an 
ensemble specified by  contact number  $n_f$ with the polymer distal end
$i=N$ being in a contact  position (as indicated by the superscript ``$c$'').

Finally, we introduce the partition function for a completely unfolded polymer
segment running from sequence index $j$ to sequence index $i$, weighted by
additional terms on the two end residues. Explicitly, the path
integral of the unfolded segment takes the form
\begin{eqnarray}\label{M1}
M_{i,j}(\mu_1,\mu_2)
&=&\prod_{k=j}^i\int\left[\sqrt{{\beta\kappa\over2\pi}}\right]^d
d^dx_k\left[1-\Delta_k\right]\\
&&\hspace{.1cm}\times
e^{-{\beta\kappa\over2}\sum_{s=j+1}^{i}|\nabla\vec{x}_s|^2
-{\beta\kappa\over2\mu_1}|\vec{x}_i|^2
-{\beta\kappa\over2\mu_2}|\vec{x}_j|^2} \nonumber
\end{eqnarray}

\begin{figure}
\centerline
{\epsfxsize = 7.0cm \epsffile{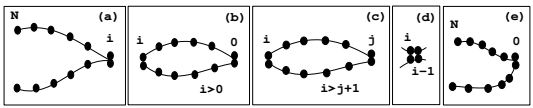}}
\vspace{.3cm}
\caption{
Using the $M_{i,j}(\mu_1,\mu_2)$ functional to represent the path integral of
several diagrams. (a) For a polymer segment starting from an 
``open'' distal end $N$ and ended at a ``contact'' residue $i>0$, without any 
residue between $N$ and $i$ being in contact position, the path integral is 
equal to $M_{N,i+1}(\infty,{1\over1+\gamma})$. Here the contact 
energy gain from residue $i$ is not included. 
(b) For a polymer segment staring from the proximal end $0$ and
ended at a ``contact'' residue $i$, without any residue between $N$ and $0$
being in contact position, the path integral reads 
 $M_{i-1,1}({1\over1+\gamma},1)$. (c) For polymer segment starting
from one contact residue $i$ to another one $j<(i-1)$, without any residue
between $i$ and $j$ being in contact position, the path integral is 
$M_{i-1,j+1}({1\over1+\gamma}, {1\over1+\gamma})$. (d) For
a polymer segment only containing two contiguous contact residues, the path 
integral reduces to 1.
(e) If the entire polymer is ``open'', i.e., none of the residues
are in contact position, the path integral equals $M_{N,1}(\infty,1)$.
}\label{diagram}
\end{figure}

In Fig.\ref{diagram}, 
we show how a variety of path integrals, which form the building
blocks for the entire partition function, can be represented in terms of
$M_{i,j}(\mu_1,\mu_2)$. This notion allows to break up the partition
function into a sum of partition functions for smaller subsystems. 
Specifically, we have for $n_f>0 $,
\begin{equation}
Z_{N,n_f}=Z_{N,n_f}^{c}
+\sum_{k=n_f}^{N-1}M_{N,k+1}(\infty,{1\over 1+\gamma})Z_{k,n_f}^{c}
\label{ZN1}
\end{equation}
and for $n_f >1$,
\begin{eqnarray}
Z_{N,n_f}^{c}&=&qe^{\beta V_1}\Biggl\{e^{\beta V_2}Z_{N-1,n_f-1}^{c}
\label{ZN2}\\
&&\hspace{.1cm}
+\sum_{k=n_f-1}^{N-2}M_{N-1,k+1}({1\over 1+\gamma},{1\over 1+\gamma})
Z_{k,n_f-1}^{c}\Biggr\}
\nonumber
\end{eqnarray}
with 
 $q= 
 {4\pi\over3}r_0^3\left[\sqrt{{\beta\kappa\over2\pi}}\right]^3\ll1$.
These are supplemented by the  boundary conditions $M_{i,j}(\mu_1,\mu_2)=0$ 
if $j>i$,
 $Z_{N,0}=M_{N,1}(\infty,1)$ (the coil state with $n_f=0$), 
and 
 $Z_{N,1}^{c}=qe^{\beta V_1}\left[M_{N-1,1}({1\over1+\gamma},1)
 +e^{\beta V_2}\delta_{N,1}\right]$. Note that it is straightforward to obtain
 $Z_{N,N}=[qe^{\beta(V_1+V_2)}]^N$; also, $Z_{N,0}$ is always less than
unity. These will be used later. 

The final step of our solution involves finding a recursive formula for 
$M_{i,j}(\mu_1,\mu_2)$. Recall that we have assumed 
that the contact distance $r_0$ is far smaller than the 
thermodynamic inter-residue distance $1/\sqrt{\beta\kappa}$ and that 
therefore we can replace the Gaussian connectivity 
 ${\beta\kappa\over2}|\nabla\vec{x}_k|^2$ by
 ${\beta\kappa\over2}|\vec{x}_k|^2$ if $\vec{x}_{k-1}$ is in contact position.
If we consider a particular $i,j$ pair, we can integrate out the coordinate
of the $i^{\mbox{th}}$ residue; this will yield the two terms
\begin{eqnarray}\label{M2a}
&&M_{i,j}(\mu_1,\mu_2) \\
&&\hspace{.1cm}
=\left[{\mu_1\over1+\mu_1}\right]^{3\over2} M_{i-1,j}(\mu_1+1,\mu_2)
-qM_{i-1,j}(1,\mu_2)\nonumber
\end{eqnarray}
The terms correspond to whether this residue is not or is in contact. 
Again, this must be supplemented by the boundary condition
\begin{equation}\label{M2}
M_{j,j}(\mu_1,\mu_2) = [\mu_1 \mu_2/(\mu_1 +\mu_2)]^{3\over2}-q
\end{equation}

\begin{figure}
\centerline
{\epsfxsize = 7.5cm \epsffile{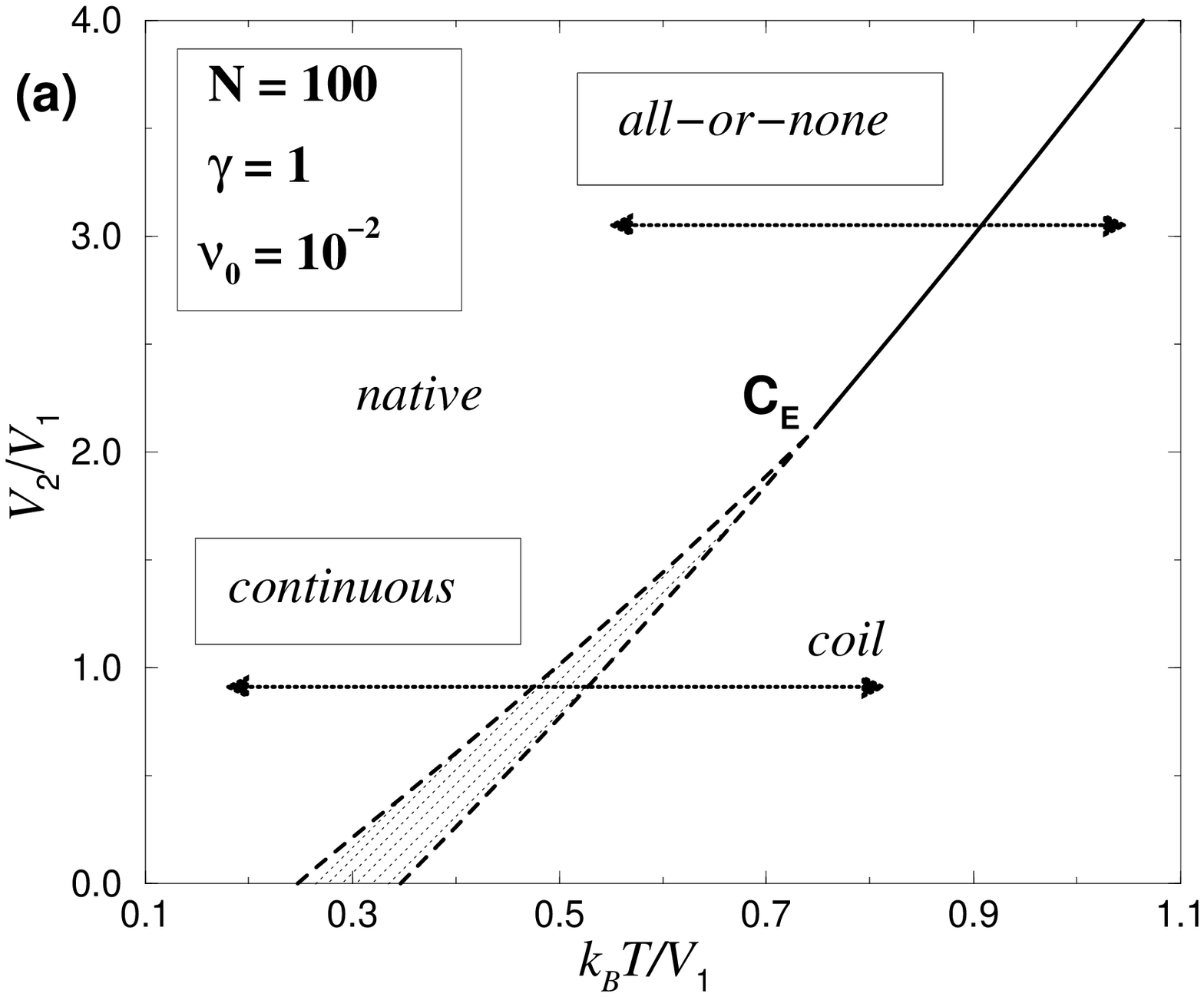}}
\centerline
{\epsfxsize = 7.5cm \epsffile{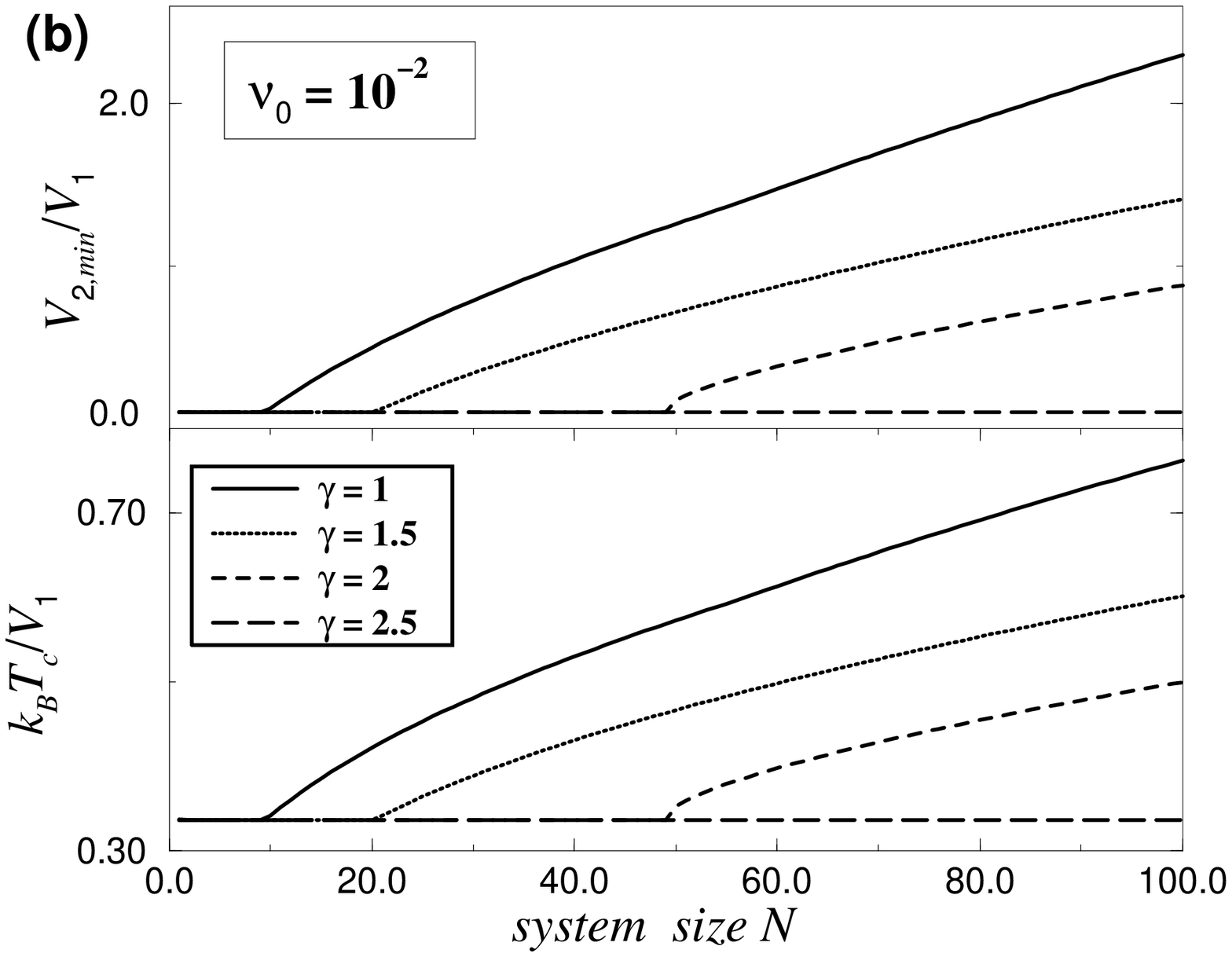}}
\caption{ 
(a): The phase diagram for a $N=100$, $\gamma=1$ $\beta$-hairpin. Here
the side chain packing energy $V_2$ and temperature $k_BT$ is scaled in the 
unit of the hydrogen bond energy $V_1$.
Above the point $\rm C_E$, the polymer has a 2-state all-or-not transition.
Below $\rm C_E$ is the ``intermediate'' regime (the shadow enclosed by the 
dashed line) where transition might be continuous. As $\gamma$ increases, 
$\rm C_E$ decreases toward zero and the intermediate regime eventually 
vanishes.
(b): The minimal side chain packing energy $V_{2,min}$ required for
an all-or-none 2-state folding transition, on the variation of 
backbone stiffness modulation $\gamma$ and hairpin size $N$ ($T_c$ at 
 $\rm C_E$ is also computed). 
}\label{phase}
\end{figure}

Using these recursion relations, 
we now can compute the thermodynamic probability 
of the native ($n_f =N$), unfolded coil ($n_f=0$), and partially folded 
ensembles.  The partition function is dominated by the highest probability 
state.
An all-or-none transition will occur if this dominant state changes from coil 
to native (as the temperature is lowered) without passing through an
extended region of parameter space in which intermediate states dominate; 
otherwise, the transition will be continuous.  Our results indicate that at a 
fixed value of the (cooperative stiffness increase) $\gamma$  the transition 
between coil and
native states is 2-state-like as long as ${V_2\over V_1}$ is above some 
point $\rm C_E$. Here $\rm C_E$ is the triple phase coexisting point 
(the coil, native, and one partially folded ensembles).  This is shown in 
 fig.\ref{phase}(a) for the case of 
 $\nu_0={4\pi\over3}\left[
 {r_0\sqrt{\kappa}\over\sqrt{2\pi V_1}}\right]^3 =.01$.  

As $\gamma$ increases, the intermediate regime shrinks and eventually 
disappears.
In other words, in order to obtain an all-or-none transition, we must have a 
minimal side chain packing strength $V_{2,min}$ with respect to a particular 
set of $V_1$ and $\gamma$; for large enough $\gamma$, $V_{2,min} =0$. This 
behavior is plotted in  fig.\ref{phase}(b). 

Although it is not relevant for the biological system, it is interesting 
from the general statistical physics perspective to consider what
happens to the minimal $V_2$ as $N$ gets large.
If we define the folding temperature  at the point $\rm C_E$ as 
$T_c=1/k_B\beta_c$, then  $V_{2,min}$ and $\beta_c$ satisfy 
\begin{eqnarray*}
Z_{N,0}\bigr|_{\beta_c}=Z_{N,N}\bigr|_{\beta_c}=\mbox{max}
 \left\{Z_{N,1},Z_{N,2},\dots,Z_{N,N-1}\right\}\bigr|_{\beta_c}
\end{eqnarray*}
where ``$\mbox{max}\left\{x_1,x_2,\dots,\right\}$'' 
picks up the maximal one in the set $\{x_1,x_2\dots\}$. This leads to
 $Z_{N,0}|_{\beta_c}=Z_{N,N}|_{\beta_c}\ge Z_{N,N-1}|_{\beta_c}$. 
From the recursion relation (\ref{ZN1}), (\ref{ZN2}), 
we find that the ``large $N$'' components in $Z_{N,N-1}$ is 
 $(N-2)[qe^{\beta V_1}]^{N-1}e^{(N-2)\beta V_2}
 M_{k,k}({1\over1+\gamma},{1\over1+\gamma})$ 
with the one-residue loop entropy $M_{k,k}$ described by eqn.(\ref{M2a}).
Combining this with the previous formula for $Z_{N,N}$, the
above inequality requires
\begin{eqnarray}
1 & \ge & q_ce^{\beta_c[V_1+V_{2,min}]} \\
 & \gtrsim &
 (N-2)M_{k,k}({1\over1+\gamma},{1\over1+\gamma}) e^{-\beta_cV_{2,min}} \nonumber
\end{eqnarray}
with  $q_c\approx{4\pi\over3}r_0^3
\left[\beta_c\kappa/2\pi\right]^{3\over2}\ll 1$.
Clearly, this implies that at large $N$ limit, the surface tension 
penalty (arising from $\gamma$ and $V_2$) must be large enough to compete 
with the 
combinatory entropy effect.
At small $\gamma$, the only possible behavior consistent with the
above inequality is $T_c\sim N^{2/3}$ and $V_{2,min}\sim N^{2/3}\ln N$.
If $\gamma$ is large, it appears that there is another choice, namely
the vanishing of the one-residue entropy,  
$M_{k,k}({1\over1+\gamma},{1\over1+\gamma})\sim O(1/N)$. In this case, we find that $T_c$ 
approaches a constant  
$k_BT_c\approx 0.83r_0^2(1+\gamma)\kappa+O(1/N)$ and
$V_{2,min}\approx k_BT_c\ln[(1+\gamma)^{3/2}-1]]-V_1+O(1/N)$. One
should note however, that in this limit our original assumption
that one can approximate factors of the form $\exp{\gamma\kappa r_0^2}$
as unity is no longer accurate, and the model crosses back
to the requirement of an $N$ dependent cooperative interaction. 

To summarize, we find that 2-state behavior arising from short-ranged 
interactions can only exist in finite size hairpin systems. For
such a finite system, increasing the stiffness of the bound parts of the 
chain will lead to this behavior at biologically relevant values
of the chain length $N$. These results are  
consistent with those obtained in other two-chain models 
\cite{Cule,Dauxois1}. 

\section{The Folding/Unfolding Rate}

We now compute the unfold/folding rate for side chain packing strength 
 $V_2\ge V_{2,min}$ (i.e., the parameter regime above the triple-phase point 
 $\rm C_E$). The transition is  2-state-like, and the transition  rate
is of the Arrhenius form 
 $k_{tx}\sim k_0 e^{-\beta\Delta F}$, 
where $k_0$ is a kinetic prefactor (which might be different for 
folding versus unfolding transitions), and the exponent 
$\Delta F$ is the free energy difference between 
the saddle point configuration (the transition state structure) and the 
metastable state. In our system, the Arrhenius form 
is just the thermodynamic probability ratio between the transition ($tx$) and 
metastable ensembles, $Z_{tx}/Z_{meta}$; here,
$Z_{meta}=Z_{N,N}$ if $T>T_f$, 
$Z_{meta}=Z_{N,0}$ if $T<T_f$  and $T_f$ is the folding temperature defined by 
$Z_{N,0}\bigr|_{T_f}=Z_{N,N}\bigr|_{T_f}$.

Since there are $3N$ degree of freedom in our model, in general one can 
expect the existence of multiple saddle point configurations. Each 
configuration specifies a particular ``pathway'' towards folding/unfolding. 
The saddle point configurations must be partially folded states and their 
contact residues could be inhomogeneously distributed. This inhomogeneity and 
the number of contacts in turn determine the thermodynamic probabilities of 
these configurations. Among them, the most likely pathway is mediated by the 
configuration that has the maximal thermodynamic probability. Due to the 
surface tension effect (arising via nonzero $\gamma$), only two 
particular structures need be considered, i.e., the polymer zipping from 
either the $\beta$ turn or the distal end. As suggested by 
experiment \cite{Munoz2,Munoz1} and by the simple logic
that $\vec{x}_0 =0$ biases the polymer towards folding, 
the configuration that the ``folding''
droplet emerges from the $\beta$ turn is the most likely one (fig.\ref{h1}(b)).
If we define the restricted partition for this droplet (i.e., the polymer 
zipping from $\beta$ turn to sequence index $i$, $0<i<N$) as $Y_i$, from
the recursive relation, we have
\begin{equation}\label{YS}
Y_k=\left[qe^{\beta[V_1+V_2]}\right]^kM_{N,k+1}(\infty,{1\over 1+\gamma})
\end{equation}
The transition state, therefore, is defined as the particular state 
 $Z_{tx}=Y_k$, with $Y_k$ satisfying $Y_k\le Y_{k\pm1}$. 
\begin{figure}
\vbox
{\centerline
{\epsfxsize = 7.2cm \epsffile{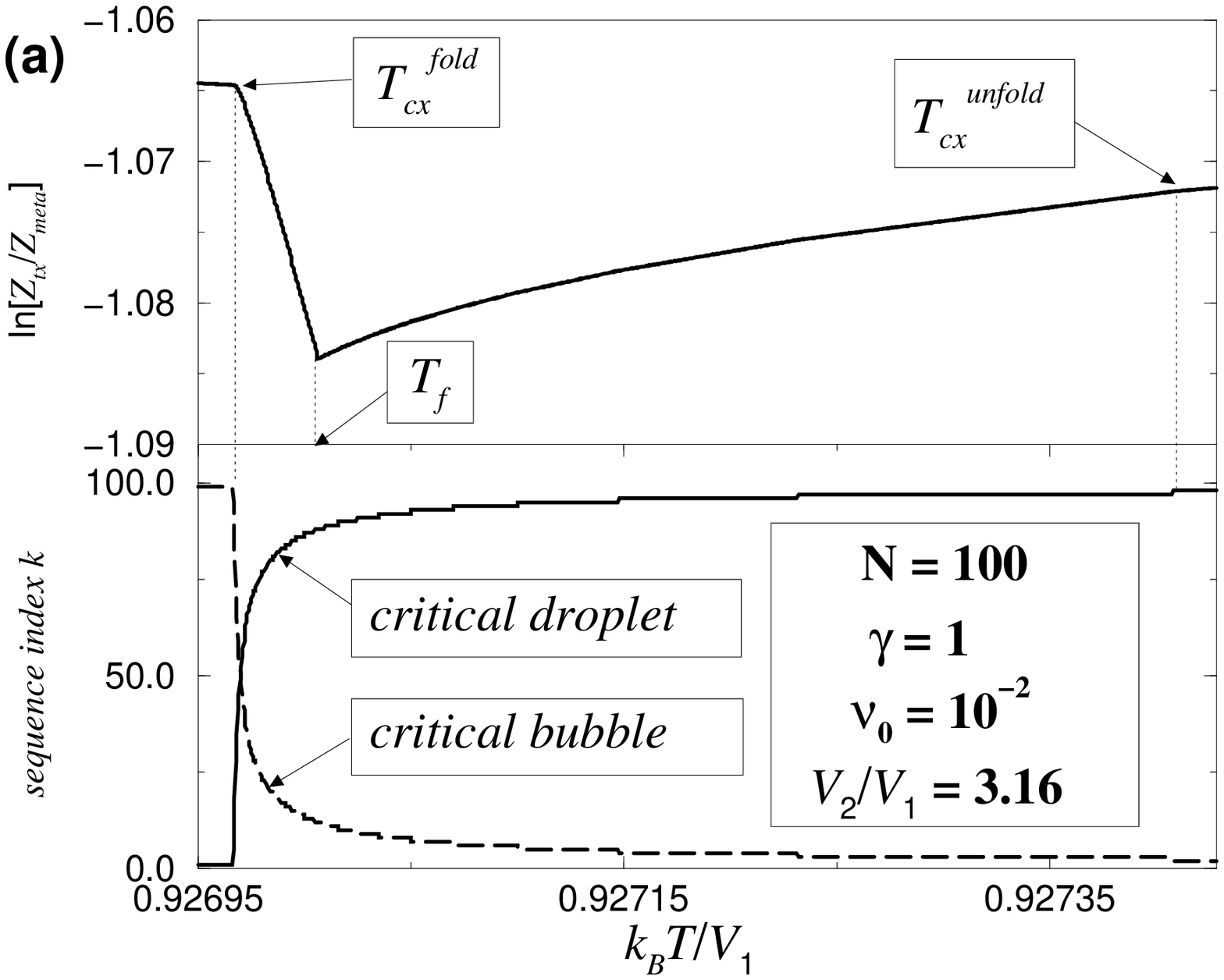}}
\centerline
{\epsfxsize = 7.2cm \epsffile{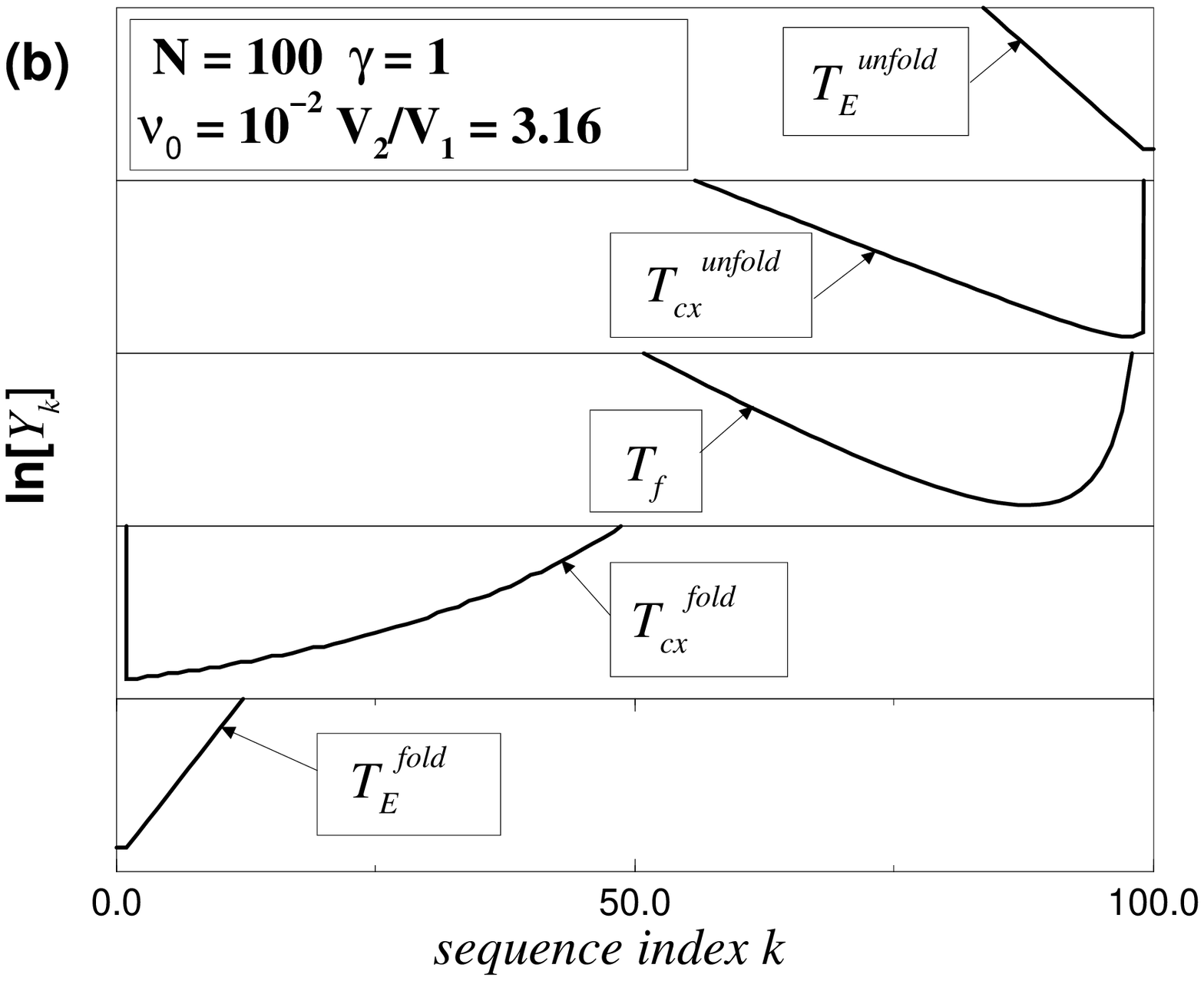}}
}
\caption{
(a): The Arrhenius component $\ln[Z_{tx}/Z_{meta}]$ for
folding/unfolding transition and the corresponding critical droplet/bubble
size on the variation of temperature.
Note the ``crossover'' regime ($T>T_{cx}^{unfold}$ and $T<T_{cx}^{fold}$).
(b): The appearance of crossover temperature $T_{cx}$ and end temperature 
 $T_E$ is determined by the critical droplet/bubble size. When
 $T=T_E^{unfold}$ (the top one), $Z_{N,N}=Y_{99}$. When
 $T=T_{cx}^{unfold}$ (the second top one), $Y_{99}=Y_{98}$.
Likewise, when $T=T_{cx}^{fold}$ (the second lowest one), 
 $Y_2=Y_1$, and when $T=T_E^{fold}$ (the lowest one), 
 $Y_1=Z_{N,0}$. The transition state is the lowest
 $\{\ln Y_k\}$ on every curve.
}\label{transition}
\end{figure}

While computing the transition state, we find that for a particular set of
the parameters ($V_2$, $V_1$, $\gamma$, $r_0$), the behavior of the folding
droplet or the unfolding bubble is characterized by several temperatures: 
 $T_E^{unfold/fold}$, $T_{cx}^{unfold/fold}$, and
 $T_f$ (fig.\ref{transition}). 
For $T>T_E^{unfold}$ (or $T<T_E^{fold}$), there is no 
transition state and the transition from native (coil) to coil (native) state 
is purely relaxational. Here the ``end'' temperature 
$T_E^{unfold}$ is defined by 
 $Z_{N,N}\bigr|_{T_E^{unfold}}=Y_{N-1}\bigr|_{T_E^{unfold}}$, and 
 $T_E^{fold}$ is defined by 
 $Z_{N,0}\bigr|_{T_E^{fold}}=Y_1\bigr|_{T_E^{fold}}$.
For $T_{cx}^{unfold}<T<T_E^{unfold}$ (or $T_{cx}^{fold}>T>T_E^{fold}$), on the
other hand, the size of the ``critical'' unfolding 
bubble (folding droplet) is one. Here the ``crossover'' temperature 
 $T_{cx}^{unfold}$ is defined by 
 $Y_{N-2}\bigr|_{T_{cx}^{unfold}}=Y_{N-1}\bigr|_{T_{cx}^{unfold}}$, and
 $T_{cx}^{fold}$ is defined by 
 $Y_{2}\bigr|_{T_{cx}^{fold}}=Y_{1}\bigr|_{T_{cx}^{fold}}$.
Finally, for $T_{cx}^{fold}<T<T_{cx}^{unfold}$, the critical unfolding bubble
(if $T>T_f$) or folding droplet (if $T<T_f$) has a size $k$ between 1 and 
 $N-1$, determined by variational method, $Y_k\le Y_{k\pm1}$.

For completeness, we also computed the folding and unfolding rates based on 
the configuration that the droplet is initiated from the distal end instead of 
the $\beta$ turn. It turns out that its folding/unfolding
rate is $10^3$-fold lower than the previous configuration. 
This confirms our assumption and is consistent with the experimental 
observation that additional inter-chain interaction near the $\beta$ turn 
enhances the folding rate \cite{Munoz2,Munoz1}. 
Thus we conclude that the dominant pathway for polymer folding is zippered 
from the $\beta$ turn; whereas the pathway for unfolding is un-zippered from 
the distal end.

\section{Discussion}

We have analyzed a simple $\beta$ hairpin model and discussed the significance
of side chain packing regularities. We found that side-chain packing 
regularities are necessary to generate 2-state transitions between coil and
native structures. There will be an upper limit to the hairpin
size $N$ for which this behavior persists but this does not
affect its applicability to biologically relevant
finite-sized chains. Since it has
been shown that folding of a $\beta$-hairpin captures much of the basic
physics of protein folding\cite{Munoz2} (imagine the folding of two
 $\alpha$-helices connected by a $\beta$ turn; the coupling between the
``secondary'' and ``tertiary'' structure there is similar to the coupling 
between ``primary'' backbone stiffness and ``secondary'' contact 
formation here), our model can provide a fundamental understanding of how a 
protein can fold.

Why is the coupling induced by side chain packing important? The reason is as
follows. In general, the ensemble space of a hairpin polymer
contains two entropic components: one is configurational regarding
local loop entropy and another is combinatory indicating the total
possible arrangement of the hydrogen bonds in a partially folded
state. The combinatory part grows exponentially as the system size
increases, for states which have a finite fraction of the possible
hydrogen bonds. This might compensate the configurational entropy
loss compared to the coil state and the relatively lower
hydrogen-bond energy gain compared to the native hairpin state. The
partially folded state then becomes thermodynamically predominant, and
the 2-state transition between coil and native states will be
destroyed. To avoid this situation, a collective effect must be
imposed. For the on-lattice model, this needed effect can come from 
the restricted arrangement of the polymer residues, since the alignment of
one part of the polymer will affect the others and the influence 
ultimately reach the whole length of the chain \cite{Chen}. On the contrary, 
there is no such effect in off-lattice models; instead, one has to design the
Hamiltonian carefully to obtain the desired  2-state behavior. 

It appears that the ``side-chain packing'' regularity is the essential 
ingredient to allow the all-or-none folding transition 
\cite{Munoz2,Munoz1,Chen}; this is perhaps similar to the ideas of
``hydrophobic collapse''
and  ``non-additive'' force put forth by other groups \cite{Klimov,Peter}. 
The side-chain packing is essentially 
dependent on the ``matching'' of the local peptide backbone conformation and
the hydrogen-bond formation \cite{Munoz1}. We have argued that
this fact leads to new cooperative terms in the Hamiltonian,
as having some residues hydrogen-bonded will bias other residues
to form their native contacts and also locally restrict conformational
entropy. One critical consequence of this is the creation of an
effective ``surface tension'' between the folded and unfolded
regimes. This energy cost will compete with
the combinatory entropy gain of a partially folded state,
and the all-or-none transition can be restored. This side-chain
effect is not present in systems undergoing the $\alpha$ helix-coil 
transition, manifesting the essential difference between $\alpha$-helix
and $\beta$-hairpin formation.

Our model is, of course, greatly simplified compared to the 
actual $\beta$-hairpin system. One immediate criticism
is that we have neglected non-native 
hydrogen bond formation by using the $\rm G\bar{o}$ interaction.
To date, NMR evidence suggests \cite{Munoz1} that 
states with non-native bonds have minimal thermodynamic weight, lending
support to the adequacy of this approach. These neglected effects, however, 
might produce local minima in the energy landscape and trap misfolded 
structures, leading to a glassy molten globule. 

Another problem is our use of uniform strengths for all interactions.
This was necessary because of our desire to develop a recursion relationship
which enables us to do calculations for reasonably high values of $N$.
One can extend the model to heterogeneous couplings, but then
one will have to resort to exact evaluation of each of the
$2^N$ different states of the $\rm G\bar{o}$ contacts. This would
limit us to small polymers; also, the cooperative ``zippering'' behavior in 
folding transition might be altered to start from the center of a hydrophobic 
cluster instead of the $\beta$ turn \cite{Klimov}. In any case, we do not believe 
that modest heterogeneity will lead to any significant changes in our results
regarding the source of the cooperativity in this class of systems.

C. Guo gratefully thanks Margaret Cheung, Ralf Bundschuh $\&$ Terence Hwa
for useful discussions. H. Levine acknowledges the 
support of the US NSF under grant DMR98-5735.

\end{document}